\documentclass[preprint2]{aastex631}

\begin{document}


\title{Confirmation of sub-solar metallicity for WASP-77Ab from JWST thermal emission spectroscopy}

\author[0000-0003-3829-8554]{Prune C. August}
\affiliation{Department of Astronomy \& Astrophysics, University of Chicago, Chicago, IL, USA}
\affiliation{Institute of Physics, École polytechnique fédérale de Lausanne (EPFL) Lausanne, Switzerland}

\author[0000-0003-4733-6532]{Jacob L.\ Bean}
\affiliation{Department of Astronomy \& Astrophysics, University of Chicago, Chicago, IL, USA}

\author{Michael Zhang}
\altaffiliation{51 Pegasi b Fellow}
\affiliation{Department of Astronomy \& Astrophysics, University of Chicago, Chicago, IL, USA}

\author{Jonathan Lunine}
\affiliation{Department of Astronomy, Cornell University, Ithaca, NY, USA}

\author[0000-0002-6215-5425]{Qiao Xue}
\affiliation{Department of Astronomy \& Astrophysics, University of Chicago, Chicago, IL, USA}

\author{Michael Line}
\affiliation{School of Earth and Space Exploration, Arizona State University, Tempe, AZ, USA}

\author[0000-0002-9946-5259]{Peter C. B. Smith}
\affiliation{School of Earth and Space Exploration, Arizona State University, Tempe, AZ, USA}




\begin{abstract}
We present the dayside thermal emission spectrum of WASP-77Ab from 2.8 -- 5.2\,$\mu$m as observed with the NIRSpec instrument on the James Webb Space Telescope (JWST). WASP-77Ab was previously found to have a sub-solar metallicity and a solar carbon-to-oxygen (C/O) ratio from H$_2$O and CO absorption lines detected using high-resolution spectroscopy. By performing atmospheric retrievals on the JWST spectrum assuming chemical equilibrium, we find a sub-solar metallicity [M/H]=$-0.91^{+0.24}_{-0.16}$ and C/O ratio $0.36^{+0.10}_{-0.09}$. We identify H$_2$O and CO and constrain their abundances, and we find no CO$_2$ in the spectrum. The JWST and high-resolution spectroscopy results agree within $\sim1\sigma$ for the metallicity and within 1.8$\sigma$ for the C/O ratio. However, our results fit less well in the picture painted by the shorter wavelength spectrum measured by HST WFC3. Comparing the JWST thermal emission spectra of WASP-77Ab and HD\,149026b shows that both hot Jupiters have nearly identical brightness temperatures in the near-infrared, but distinctly different atmospheric compositions. Our results reaffirm high-resolution spectroscopy as a powerful and reliable method to measure molecular abundances. Our results also highlight the incredible diversity of hot Jupiter atmospheric compositions.


\end{abstract}



\section{Introduction}\label{sec:intro}
Atmospheric metallicity and carbon-to-oxygen (C/O) ratio are now well-established tracers for exoplanet characterisation \citep{Madhusudhan_2012, Fortney_2013, Venturini_2016}. In particular, their relevance regarding planetary formation processes has been widely discussed in the literature \citep{1996Icar..124...62P, Mordasini_2016, madhu_review}. The James Webb Space Telescope (JWST) marks an important step in the field because it has the sensitivity to obtain high precision data over a broad range of wavelengths for a wide range of exoplanets, and therefore allows for strong and reliable constraints on these key markers \citep{Greene2016}.

Hot Jupiters are giant gaseous bodies transiting very close to their host stars. These highly irradiated planets are unlike any of our Solar System planets, and questions like their origins, as well as how they tie into a bigger picture of planet formation, still remain largely unanswered \citep{Fortney_2021}. Together with their obvious observational advantages, they thus make excellent candidates for transit spectroscopy and specifically atmospheric characterisation.

WASP-77Ab \citep{maxted13} is a great target for such endeavours: this hot Jupiter is one of the highest signal-to-noise (S/N) planets for thermal emission measurements in the near-infrared \citep{Kempton_2018}\footnote{See tabulated values for known planets at the TESS Atmospheric Characterization Working Group webpage: \url{https://tess.mit.edu/science/tess-acwg/}.}.
In a breakthrough result, observations using ground-based high-resolution spectroscopy (HRS) with the Immersion GRating INfrared Spectrometer (IGRINS) constrained the metallicity to be solar and C/O ratio to be solar based on the detection of H$_2$O and CO lines \citep{line2021solar}.

Subsequent Hubble and Spitzer Space Telescope data confirmed the presence of water vapour on WASP-77Ab's dayside \citep{mansfield2022confirmation}, which indicates C/O values $<$ 1 \citep{kreidberg2015, benneke2015}.  However, there are lingering questions about the composition of the planet because the Hubble data had some systematic outliers that couldn't be fit by 1D models and because the retrieved metallicities differed somewhat (albeit only at the 1.8$\sigma$ confidence level), with \citet{mansfield2022confirmation} estimating a solar to potentially 3x super-solar metallicity.

The low metallicity for WASP-77Ab inferred by \citet{line2021solar} is surprising from an origins standpoint because gas giants are expected to accrete planetesimals that pollute their growing atmospheres during formation. Nevertheless, recent work suggests that subsolar atmospheric abundances for close-in giant planets indicates formation beyond the CO$_2$ snowline \cite[$a >$ 5\,au,][]{bitsch22}. The results for WASP-77Ab are also surprising given the recent series of strong CO$_2$ detections in hot Jupiter atmospheres that are indicative of enhanced atmospheric metallicities \citep{2023Natur.614..649J, 2023Natur.614..659R, 2023Natur.614..664A, HD149paper}. Because of this and the very different data processing and analysis needed for ground- vs.\ space-based data it is important to provide an independent and precise measurement of these quantities for WASP-77Ab.

In this article, we analyze the dayside emission spectrum of WASP-77Ab obtained using the NIRSpec instrument on JWST. NIRSpec is an ideal instrument for this project because its bandpass covers key carbon- and oxygen-bearing molecules like H$_2$O, CO, CO$_2$ and CH$_4$. NIRSpec is also well suited for studying WASP-77Ab because it has a slit (1.6\arcsec x 1.6\arcsec) that allows the rejection of contaminating light without requiring roll angle constraints. The diffraction-limited point spread function of NIRSpec is significantly smaller (FWHM $\approx$ 0.17\arcsec) than the separation  of the two stars \citep[$\sim$3\arcsec, $\Delta{\textrm{mag}} \sim$ 2;][]{maxted13}, ensuring minimal blending and allowing for precise characterization of WASP-77Ab.
We infer fundamental diagnostics like the atmospheric metallicity and C/O ratio, as well as estimates of the thermal structure of the atmosphere and its chemistry. We place our results in the context of the previous estimates provided by \citet{line2021solar} and \citet{mansfield2022confirmation}, and we provide support for HRS as a method to retrieve chemical abundance constraints.

We describe how we obtained WASP-77Ab's dayside emission spectrum in Section \ref{sec:obs}, with a focus on the partial phase curve detection. Section \ref{sec:retrievals} is dedicated to the atmospheric retrievals and their results. We also provide a comparison with other data sets for WASP-77Ab. To further illustrate the diversity of hot Jupiters, we contrast WASP-77Ab with HD\,149026b, observed in eclipse by JWST NIRCam (\citealt{HD149paper}), in Section \ref{sec:planeto}.
Finally, we summarize our results in Section \ref{sec:ccl}.

\begin{figure*}
    \includegraphics[scale=0.97]{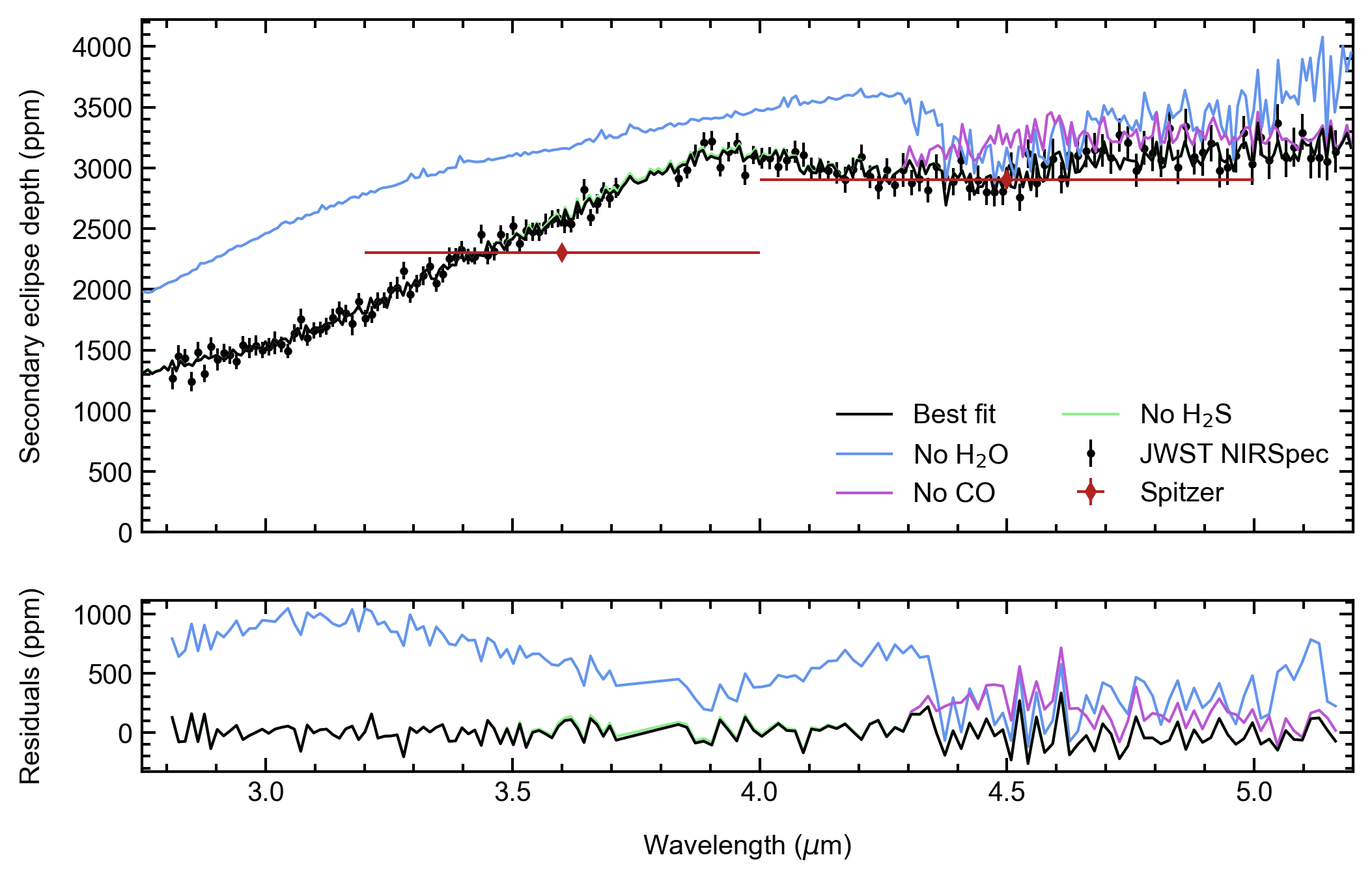}
    \caption{Thermal emission spectrum of WASP-77Ab as observed by JWST NIRSpec/G395H, along with the best fit model (black line). The different coloured lines correspond to best fit spectrum computations with different molecular opacities zeroed out. The bottom panel shows the residuals.}
    \label{fig:spectrum}
\end{figure*}

\section{Observations and data reduction}\label{sec:obs}
We observed a secondary eclipse of WASP-77Ab on August 24-25, 2022 between 22:12 - 04:36 UTC with JWST NIRSpec (program GTO 1274, J.~Lunine PI). The observations used the Bright Object Time-Series (BOTS) mode, with the G395H grating and F290LP filter combination \citep{filtercomb}. The duration of this observation was 6.4 hours for a total of 1419 integrations, and 17 groups per integration. The observation began 2.44 hours before the secondary eclipse and continued for 1.80 hour after eclipse egress. The wavelength coverage for the two sensor chip assemblies (or detectors) used in this observation, NRS 1 and NRS 2, are $2.674-3.716\thinspace\mu$m and $3.827-5.173\thinspace\mu$m respectively. We used the SUB32 subarray and the NRSRAPID readout pattern.

We reduced the raw data using the \texttt{Eureka!} pipeline \citep{Bell_2022}. This pipeline offers a reduction in 6 stages. The first two are essentially the same as the JWST Science Calibration Pipeline \citep[\texttt{jwst},][]{bushouse}. The following three perform background subtraction and spectral extraction, spectroscopic light curve generation, and spectroscopic light curve fitting. 

The system parameters used for the secondary eclipse are a planet orbital period P = 1.3600309 days, ratio of the planet’s semi-major axis to the host star radius a/Rs = 5.43, planetary orbital inclination = 89.4°, and planetary orbital eccentricity e = 0.0 \citep{mansfield2022confirmation}. Varying these parameters within their uncertainties did not impact the results. The secondary eclipse times were estimated by fitting the white light curves created for each of the NIRSpec detectors, and then used for all spectroscopic light curves. A nested sampling algorithm was used for all the parameter estimation. The measured secondary eclipse time is consistent with a circular orbit within the uncertainties of the most recent ephemerides \citep{IvshinaWinn2022, Kokori2022, CortesZuleta2020}. The bin size of these channels is set to $13/17$nm for NRS1/2 in an effort to balance out precision and retrieval performances. 

The data are processed into a final spectrum in Stage 6. We fine-tuned parameters such as the spectral and background aperture, outlier rejection, and binning in order to minimize the median absolute deviation of the final light curves and following the steps and recommendations of \citet{2023Natur.614..664A}. We performed tests of varying the different reduction parameters (apertures, sigma clipping and outlier rejection thresholds, binning, etc) in order to assess their impact on the resulting spectrum. All of these showed minimal to no impact on the final spectrum, indicative of its robustness vis-à-vis the data reduction process.

Except for a constant rate linear trend in time , which is a standard feature of NIRSpec time-series data \citep{2023Natur.614..659R, 2023Natur.614..664A, mikalevans23, moran23, lustigyaeger23}, the light curves are largely unaffected by systematics. In particular, there is no exponential ramp in the first few hours of the observation as has been seen for other early JWST observations \citep{HD149paper}. We therefore analyzed the full time series without trimming any data from the start.
The exceptional quality of the spectroscopic data combined with the short period of the target's orbit raised the question of whether the phase curve of the planet could be detected \citep[e.g.,][]{ersW18}. In order to investigate this, a first order phase curve model (see Equation~\ref{eq:phasecurve}) was added to the spectroscopic light curve fitting.
\begin{equation}\label{eq:phasecurve}
    f_{pc}(\phi) = 1 + A\cdot(\cos(\phi) - 1) + B\cdot\sin(\phi),
\end{equation}
where $\phi = \frac{2\pi}{P}(t-t_{sec})$ and $t_{sec}$ is the time of secondary eclipse.

The \texttt{Eureka!} partial phase curve fitting for the white light curves of each detector allowed us to retrieve relative amplitudes of $\tilde{a}_1 = 0.39 \pm 0.09$ and $\tilde{a}_2 = 0.42\pm 0.11$. The estimated phase shifts in degrees read $\phi_{1} = 18\pm10\thinspace$ and $\phi_{2} = 9\pm12\thinspace$ respectively. Taking all the values estimated through fitting the spectroscopic light curves, we found an average relative amplitude of $\mu_{a} = 0.54$ ($\sigma_{a} = 0.17$) and an average phase shift of $\mu_{\phi} = 11$° ($\sigma_{\phi} = 35$°). These results will be investigated further in a future paper looking at both the phase curve and the eclipse mapping signals in the data.

While on such a short window the phase curve detection is not statistically significant, it is physically motivated. Overall, the phase curve fitting doesn't have a determining impact on the final shape of the spectrum beyond a small shift upwards of about $\simeq 100\thinspace$ppm. Higher order systematic models such as a quadratic polynomial in time have a similar effect. 

The final resulting spectrum is shown in Figure~\ref{fig:spectrum}, along with the Spitzer datapoints collected from \citet{mansfield2022confirmation}. Our spectrum agrees with these within 2$\sigma$ for both the 3.6$\mu$m and the 4.5$\mu$m channels.  The derived error inflation factor for the spectroscopic light curves is typically $1.5$, and the typical RMS of the residuals is 230\,ppm.
From the modeling we find that we see broad absorption features of H$_2$O and CO. The spectrum doesn't show any features corresponding to a CO$_2$ signature, nor do we see any emission features indicative of a thermal inversion.

\begin{figure*}
    \includegraphics[scale = 0.97]{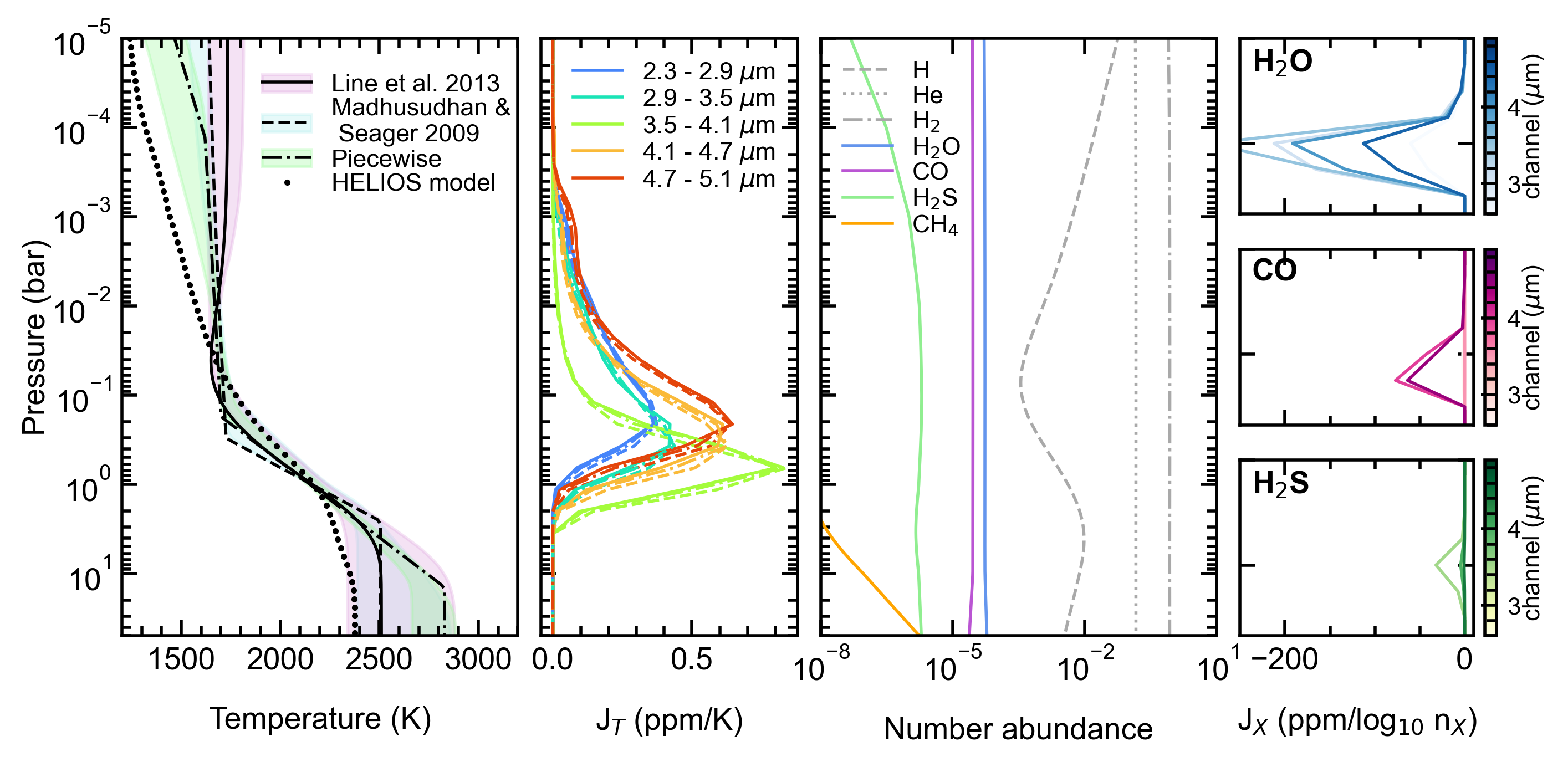}
    \caption{Description of the vertical atmospheric structure of WASP-77Ab's dayside. On the left, the first and second panel show the thermal structure along with the temperature Jacobians, which characterize the spectrum's sensitivity to small changes in the temperature. On the right, the third panel shows the chemical profiles and the three subplots investigate the abundance Jacobians of the main molecules we are sensitive to: H$_2$O, CO, and H$_2$S.}
    \label{fig:retrieval_summary}
\end{figure*}

\section{Retrievals}\label{sec:retrievals}
We performed retrievals using \texttt{PLATON} \citep{zhang19,Zhang_2020} with different temperature-pressure (TP) profile parametrizations and assumptions for the chemistry. We used the \texttt{HELIOS-K} \citep[][]{helios1, helios2} generated opacity files for the following molecules: CH$_4$, CO$_2$, CO, H$_2$O, H$_2$S, NH$_3$ and SO$_2$. We used the nested sampling algorithm, with $1,000$ live points and R=20,000 opacity files. In addition to the retrieval setup and parameters already offered by \texttt{PLATON} for the thermal and chemical structure of the atmosphere, we implemented a dilution factor as described in \citet{Taylor_2020} to account for inhomogeneities across the dayside while performing a 1-D retrieval. This parameter ranges from 0 to 1, with unity corresponding to a perfectly homogeneous dayside.

We used three models to describe the TP profile WASP-77Ab's atmosphere. The first is a version of the the analytic, level-by-level parameterization suggested by \citet{Guillot_2010}, redefined by \citet{Line_2013} to include a second irradiation stream, and implemented in \texttt{PLATON} under the name ``radiative solution''. This model has five free parameters: thermal opacity $\kappa$, visible to thermal opacity ratio of the two streams included in the model, $\gamma$ and $\gamma_2$, the percentage $\alpha$ apportioned to the second stream, and the effective albedo $\beta$. The second type of TP profile we used is a parametric model established by \citet{Madhusudhan_2009}. This model consists of placing three anchor points ($P_i, T_i$) and linking them to each other with exponential curves parameterized by $\alpha_i$. Beyond and below the first and last anchor point, the profile is assumed to be isothermal. The prior ranges for this profile are approximately the same as the ones used in \citet{line2021solar} to allow for adequate comparison. Finally, the last TP profile is a similar ``three layers'' structure with lapse rates between them, which we refer to as the piece-wise solution. There are also 6 free parameters: three pressures $P_i$ and three lapse rates $k_i$.

The chemical properties and composition of the atmosphere were probed using both equilibrium and free chemistry. For the free retrieval we modified \texttt{PLATON} to retrieve the abundances of five molecules (H$_2$O, CO, CO$_2$, CH$_4$, and H$_2$S). Their abundance profile was fixed to be constant with altitude and evaluated through a logarithmic prior ranging from $[-16,0]$ in the free retrieval. The equilibrium chemistry is built into \texttt{PLATON} via \texttt{GGChem} \citep{Woitke_2018}.

\begin{table*}
\begin{tabular}{ll||llll||l}
 Data & TP model & [M/H] & [M/H]$_{best}$ & C/O & C/O$_{best}$ & $\chi_{\nu}^{2}$ \\ \cline{1-7} 
JWST NIRSpec & L+13 & $-0.91^{+0.24}_{-0.16}$ & -0.97 & $0.36^{+0.10}_{-0.09}$ & 0.33 & 0.778 \\
 & MS09 & $-1.03_{-0.18}^{+0.34}$ & -1.16 & $0.29_{-0.06}^{+0.11}$ & 0.28 & 0.778 \\
 & Piece-wise & $-0.96_{-0.12}^{+0.16}$ & -1.04 & $0.31_{-0.05}^{+0.07}$ & 0.30 & 0.796 \\ \cline{1-7} 
JWST NIRSpec & L+13 & $-0.06_{-0.16}^{+0.13}$ & -0.00 & $0.60_{-0.05}^{+0.04}$ & 0.61 & 1.163 \\
+ HST WFC3 & MS09 & $-0.18_{-0.15}^{+0.16}$ & -0.11 & $0.47_{-0.06}^{+0.07}$ & 0.51 & 1.127 \\
 & Piece-wise & $-0.07_{-0.17}^{+0.13}$ & -0.06 & $0.52_{-0.06}^{+0.05}$ & 0.55 & 1.163
\end{tabular}
\caption{Summary of the metallicity and carbon-to-oxygen ratio estimates assuming equilibrium chemistry and} different parameterizations for the TP profile, both with and without including the HST WFC3 data. ``L+13'' refers to \citet{Line_2013}, ``MS09'' to \citet{Madhusudhan_2009} and the piece-wise model has been presented in Section~\ref{sec:retrievals}.
\label{tab:retrieval_params}
\end{table*}

Out of all the model combinations tested for the retrieval, the best fit to the spectral data, obtained with a \citet{Line_2013} style TP profile and using equilibrium chemistry, is shown in Figure~\ref{fig:spectrum}. It fits to the data with a reduced $\chi^2 = 0.78$ (150 data points and 8 free parameters), and has in particular [M/H] $= -0.97$, C/O $= 0.33$, a dilution factor $0.99$, and the TP profile shown on the far left panel in Figure~\ref{fig:retrieval_summary} (confidence intervals on the abundances are given below). The coloured lines in Figure~\ref{fig:spectrum} are the spectra computed when fixing the best fit values and zeroing the opacities of different molecules, effectively removing them from the radiative transfer but not changing the mean molecular weight of the atmosphere. The full data for this plot can be found in Table~\ref{tab:spectraldata}.

The TP profile is shown in the far left panel of Figure~\ref{fig:retrieval_summary}. In particular, the best fits estimated through three different models \citep[][and the piece-wise solution]{Line_2013, Madhusudhan_2009} are plotted in black lines along with the 2$\sigma$ confidence interval from the samples. These retrievals gave results for the composition of the atmosphere that were consistent within 1$\sigma$ with the main results \citep[using the profile from][]{Line_2013} quoted here. The retrieval outcomes of these different models is found in Table~\ref{tab:retrieval_params}.  We also compare to the TP profile calculated by HELIOS \citep{malik_2017,malik_2019}, a self-consistent forward modelling code, with a $f$ factor of 0.6 (implying close to zero albedo and zero day-to-night recirculation), a metallicity of 0.1x solar, and a C/O ratio of 0.3.  The HELIOS model predicts temperatures colder than our retrieval results at low pressures, where our retrievals are insensitive, but the temperatures are in agreement at the photospheric pressures of $\sim$30--1000 mbar.

The temperature jacobians $J_T$ in the second panel of Figure~\ref{fig:retrieval_summary} are computed by numerically deriving the spectrum $F_p$ with respect to the temperature $T$, similarly to the approach outlined by \citet{Eyre_1993} and \citet{Garand_2001}. This method, described in Equation~\ref{eq:jacT}, allows to assess how sensitive the spectral data is to small variations of temperature along the vertical pressure structure, i.e. what layers are being probed. 
\begin{equation}\label{eq:jacT}
    J_T = \frac{f_p(T + \delta T)-f_p(T - \delta T)}{2\cdot\delta T}
\end{equation}

The abundance profiles are shown in the third panel of Figure~\ref{fig:retrieval_summary}.
Similarly, one can induce small variation of the abundance of specific molecule $X$ and compute the resulting spectral fluctuations to characterise sensitivity. In this case, it is more intuitive to perform a log-derivative of the eclipse depth like in Equation~\ref{eq:jacchem}.
\begin{equation}\label{eq:jacchem}
    J_{X} = \frac{f_p(n_{X} + \delta n_{X})-f_p(n_{X} - \delta n_{X})}{2\cdot\delta n_{X}} \cdot n_{X}
\end{equation}
For the three subplots on the far right panel of Figure~\ref{fig:retrieval_summary}, we preferred a base-10 logarithm to match the x-axis graduation on the left panel.

The equilibrium chemistry model yields a metallicity of $\text{[M/H]}=-0.91^{+0.24}_{-0.16}$ (ie. about $0.1-0.2$ times solar) and a carbon-to-oxygen ratio of $\text{C/O}=0.36^{+0.10}_{-0.09}$. The free retrieval yields molecular abundances of $\log_{10}(n_{H_2O}) = - 4.26_{-0.10}^{+0.14}$ for water and $\log_{10}(n_{CO}) = - 4.58_{-0.24}^{+0.27}$ for carbon monoxide. It also puts an upper limit on the carbon dioxide abundance $\log_{10}(n_{CO_2}) < - 7.50$ at 3$\sigma$. These results compare well to the abundances estimated through equilibrium chemistry.

\begin{figure*}
    \includegraphics[scale = 0.97]{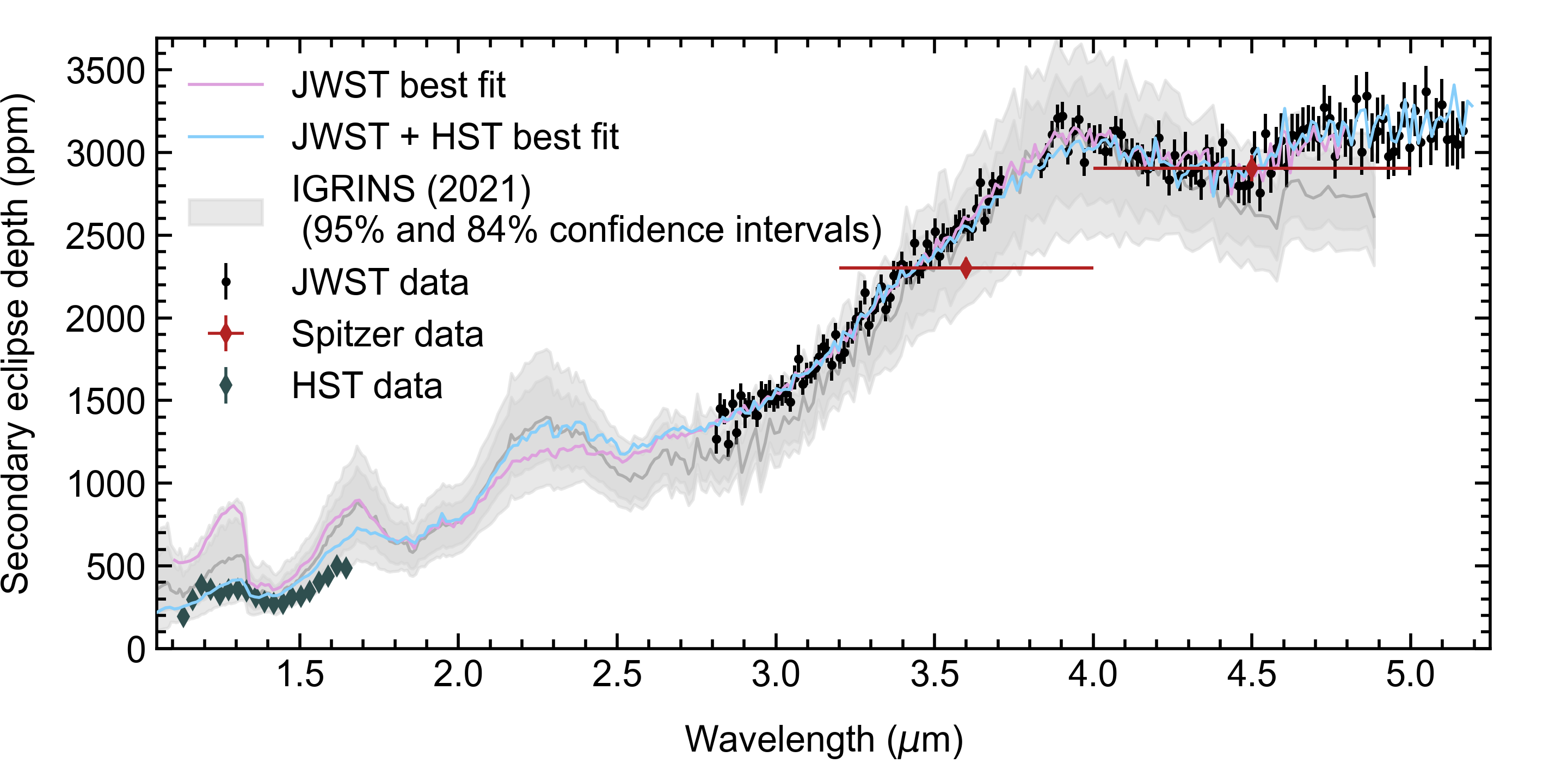}
    \caption{Concordance of our JWST NIRSpec data and retrieval results with other data sets (Spitzer, HST WFC3) and the reconstructed spectrum from high-resolution spectroscopy results with IGRINS.}
    \label{fig:datasets}
\end{figure*}

The \citet{mansfield2022confirmation} HST WFC3 data seemed to fall within the continuity of the spectrum, so we attempted a joint analysis as shown on Figure~\ref{fig:datasets}. These retrievals reevaluate metallicity and C/O ratio to higher values, [M/H] = $-0.06^{+0.13}_{-0.16}$ and C/O = $0.60^{+0.04}_{-0.05}$. However, the models do not fit the data very well. While the reduced $\chi^2$ ($1.17$, compared to the JWST-only reduced $\chi^2 = 0.78$) is not bad per se, the fit to the WFC3 data is systematically wrong by eye. The shape of the HST spectral data cannot be reproduced by our models, and this can't be fixed by adding a simple offset. Whether this is due to poor data quality, aerosols in the planet's atmosphere, or a combination is unclear, and our attempts at accounting for these in the retrievals were unsuccessful. Given the agreement between the JWST and high-resolution results, we suspect that the HST data are not fully reliable.

The JWST-only results confirm the sub-solar metallicity inferred by \citet{line2021solar} using high-resolution spectroscopy with IGRINS, agreeing at $\sim1\sigma$. The C/O ratios agree at about $1.8\sigma$, as our retrievals favour a sub-solar value. Figure~\ref{fig:datasets} shows the concordance of our spectrum and models with other datasets, namely the Spitzer and HST WFC3 datapoints as well as the \citet{line2021solar} generated spectrum from the HRS results. The PLATON retrieval results were confirmed using a retrieval with a GPU-accelerated version of \texttt{CHIMERA} \citep{Line_2013a, BrogiLine2019}, adapted for equilibrium chemistry by using \texttt{GGChem} \citep{Woitke_2018}, which yielded a metallicity of [M/H] = -0.85$^{+0.20}_{-0.12}$ and a carbon-to-oxygen ratio of C/O = $0.32^{+0.14}_{-0.09}$.

\section{Comparative planetology}\label{sec:planeto}
\begin{figure*}
    \includegraphics[scale = 0.97]{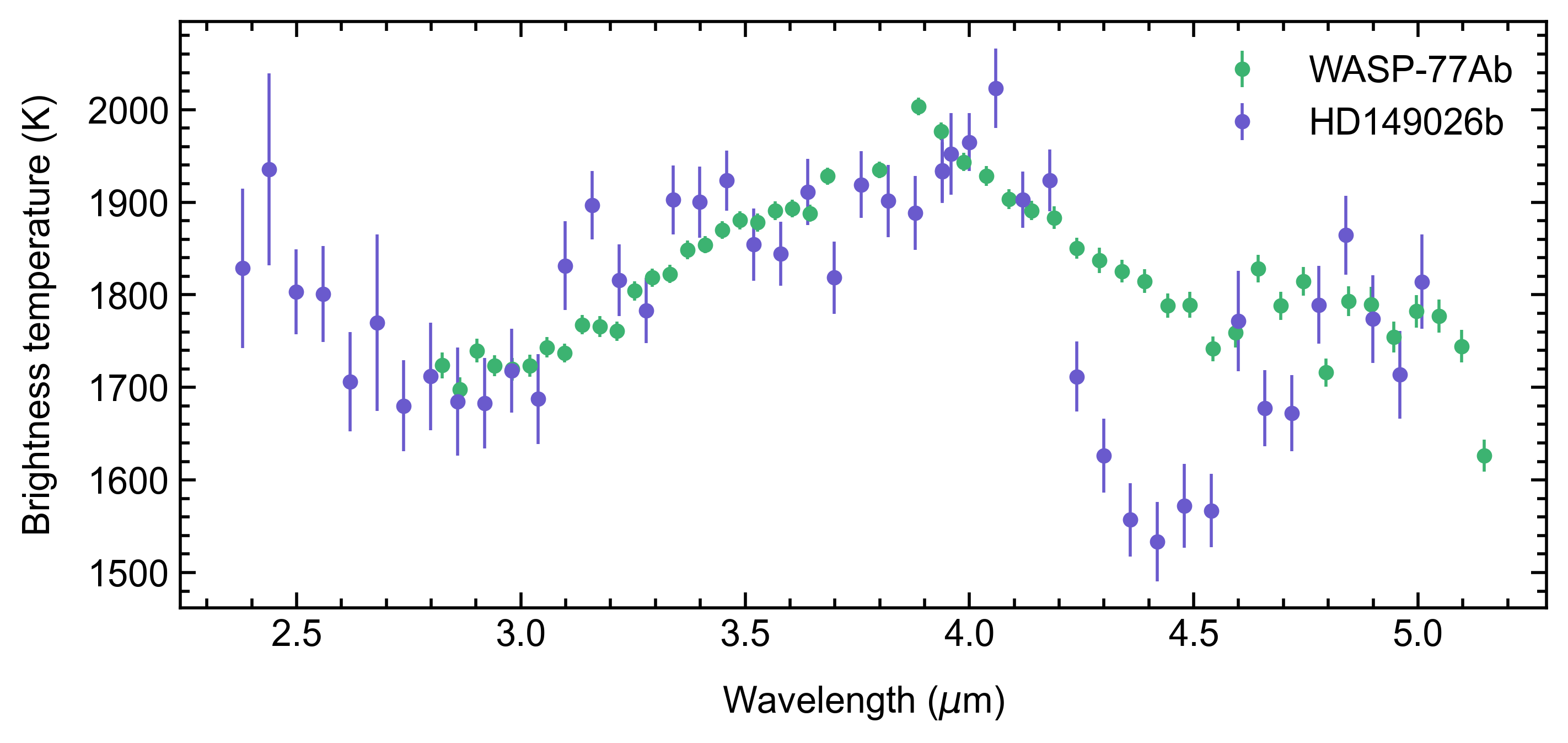}
    \caption{Planetary brightness temperatures of HD\,149026b \citep[purple points,][]{HD149paper} and WASP-77Ab (green points) from JWST thermal emission spectra. The spectrum of HD\,149026b exhibits strong absorption due to CO$_2$ that is absent from the spectrum of WASP-77Ab.}
    \label{fig:brightnesstemp}
\end{figure*}
Here we compare our spectrum of WASP-77Ab to the JWST spectrum of HD\,149026b that was recently presented by \citet{HD149paper}. These planets have similar equilibrium temperatures (1705 vs.\ 1634\,K, respectively) and now both have high-quality thermal emission spectra from 2.8 -- 5.0\,$\mu$m. To put the spectra on the same scale, we converted the measured secondary eclipse depths ($F_p/F_s$) to brightness temperatures. We used stellar spectra computed with \texttt{HELIOS} \citep{helios1, helios2} for the conversion. The stellar parameters used are $T_{\mathrm{eff}} = 5525$K, $\log\thinspace\text{g} = 4.44$, and [M/H] = 0.0 for WASP-77A \citep{Reggiani_2022}, and $T_{\mathrm{eff}} = 6085$K, $\log\thinspace\text{g} = 4.19$, and [M/H] = 0.25 for HD\,149026 \citep{HD149paper}. Then a blackbody function was fitted to the planet flux measurement in order to determine the brightness temperature at each wavelength. These temperatures are then binned by an arbitrary factor of three for clarity and shown in Figure~\ref{fig:brightnesstemp} for both planets.

Aside from the astounding S/N ratio for WASP-77Ab, it is remarkable how similar both planets' brightness temperature spectra are across the wavelength range that is dominated by H$_2$O and CO opacity. In contrast, their atmospheric chemistry is visibly different as shown by the strong CO$_2$ feature at 4.4$\mu$m in HD\,149026b. CO$_2$ is an excellent metallicity indicator for hot, giant planets \citep{lodders02,zahnle09,moses13}. Therefore, it is easily apparent that the metallicities of these two objects are very different. This illustrates the rich diversity of exoplanetary atmospheres, which was forecasted from HST WFC3 results leading up to JWST by \citet{mansfield21}.

Is the diversity of exoplanet metallicity determined by a small number of processes that relate directly to some specific aspect of formation, or is it more stochastic? \citet{line2021solar} constructed a solar system trend line of metallicity enrichment versus mass, and plotted WASP77Ab well below that line. Our metallicity result is lower but consistent at the $\sim1\sigma$ level.  They interpret the low metallicity of the atmosphere and solar C/O value in terms of a planet with most of the metals in a discrete core, and a lack of both planetesimal bombardment and significant radial migration in the disk. Our C/O value falls below the solar value, and would imply instead a radial migration that accumulates O-rich and C-poor material based on the \citet{line2021solar} systematics.  HD\,149026b, on the other hand, plots well above the solar system trend line in metallicity, has an enriched C/O value, and a massively extended (diffuse) core \citep{HD149paper}. If the latter is the result of a giant collision \citep{Ikoma_2006}, then these two objects represent outcomes of very different planetary accretion processes. This might argue that the diversity of giant planet atmospheres is reflective of a diversity in planetary formation, and stochastic events such as giant collisions are important. 

Complicating this picture is the question of whether the atmospheric compositions of planets should be interpreted in the context of an absolute scale \citep[in this paper we use the solar abundance pattern from][]{asplund09} or relative to the abundances of each of their host stars. For example, \citet{kolecki22}, \citet{polanski22}, and \citet{Reggiani_2022} have argued for the latter approach and have presented stellar abundances for WASP-77Ab (all three papers) and HD\,149026b (just the first two papers). WASP-77A is consistently found by all three studies to have solar to potentially slightly sub-solar iron abundance. However, \cite{Reggiani_2022} suggested that carbon and oxygen, which are what we are measuring in the planets and calling metallicity, were together about 2x solar. One the other hand, \citet{kolecki22} and \citet{polanski22} found the carbon and oxygen abundances more in line with the iron abundance.

For HD\,149026, all authors plus \citet{brewer16a} and \citet{brewer16b} consistently find that the iron, carbon, and oxygen are roughly 2x solar, with a slight preference for super-solar C/O from \citet{kolecki22}. If the atmospheric metallicity inferred for HD\,149026b is ``corrected'' for the host star metallicity then its atmospheric metal enhancement is reduced to $\sim$100x, which is still well above the solar system trend line \citep{kreidberg14,line2021solar,HD149paper} and significantly different than that of WASP-77Ab.

\section{Conclusion}\label{sec:ccl}
Hot Jupiters present an opportunity to measure the atmospheric abundances of both carbon- and oxygen-bearing species, which is extremely challenging for the solar system giants due to their cold temperatures \citep{guillot22}. We have presented the third thermal emission spectrum of a hot Jupiter from JWST, following WASP-18b \citep{ersW18} and HD\,149026b \citep{HD149paper}. Our spectrum of WASP-77Ab shows absorption from H$_2$O and CO, but strikingly no CO$_2$. Modeling the data indicates a strongly sub-solar metallicity for the planet's atmosphere, which is in excellent agreement with the result from the IGRINS ground-based HRS presented by \citep{line2021solar}. The agreement sets the stage for a combined analysis of these state-of-the-art data sets that may yield further insight into the properties of WASP-77Ab's atmosphere (P.~Smith in preparation).

Early results from JWST show conclusively that hot Jupiter atmospheres have diverse compositions. We have already seen that metallicities can range from sub-solar (WASP-77Ab), to solar \citep[WASP-18b:][]{ersW18}, to modestly ($\sim$10x) super-solar \citep[WASP-39b:][]{2023Natur.614..649J, 2023Natur.614..659R, feinstein23, 2023Natur.614..664A, ahrer23,tsai22}, to even very ($\sim$200x) super-solar \citep[HD149026b:][]{HD149paper}.
The diversity of giant planet atmosphere compositions that has emerged from early JWST results implies that our search for statistical trends like the mass-metallicity relationship \citep[e.g.,][]{Fortney_2013,kreidberg14} and the bulk-atmosphere metallicity correlation \citep{HD149paper} will require dozens of objects, or more.

\begin{acknowledgments}
We thank Eugenio Schisano and Ji Wang for discussions about planet host star abundances. This work is based on observations made with the NASA/ESA/CSA James Webb Space Telescope. The data were obtained from the Mikulski Archive for Space Telescopes at the Space Telescope Science Institute, which is operated by the Association of Universities for Research in Astronomy, Inc., under NASA contract NAS 5-03127 for JWST. The specific observations analyzed can be accessed via \dataset[DOI: 10.17909/3fmp-zj55]{https://doi.org/10.17909/3fmp-zj55}. JL was supported by NASA grant NNX17AL71A, and was in residence at the Dominican House of Studies, Washington DC, as the McDonald Agape Visiting Scholar, during the data analysis and preparation of the manuscript.  MZ acknowledges support from the 51 Pegasi b Fellowship funded by the Heising-Simons Foundation. 
\end{acknowledgments}

%

\vspace{5mm}
\facilities{JWST (STScI)}


\software{\texttt{Eureka!} \citep{Bell_2022}, \texttt{PLATON} \citep{Zhang_2020}, \texttt{HELIOS} \citep[][]{helios1, helios2}, \texttt{HELIOS-K}\citep{heliosk}, \texttt{CHIMERA} \citep[][]{Line_2013a, BrogiLine2019}}



\appendix
\section{Spectral data}
\begin{longtable}{lllllllll}
Wavelength & Ecl. depth & Err., low & Err., high & Best fit & No H$_2$O & No CO & No H$_2$S & No CO$_2$ \\
($\mu$m) & (ppm) & (ppm) & (ppm) & (ppm) & (ppm) & (ppm) & (ppm) & (ppm) \\ \hline
\endhead
2.808 - 2.814 & 1267 & 89 & 88 & 1394 & 2059 & 1394 & 1398 & 1394 \\
2.821 - 2.827 & 1450 & 85 & 91 & 1372 & 2090 & 1372 & 1376 & 1372 \\
2.834 - 2.840 & 1433 & 77 & 75 & 1358 & 2126 & 1358 & 1359 & 1358 \\
2.847 - 2.853 & 1237 & 79 & 80 & 1396 & 2154 & 1396 & 1398 & 1396 \\
2.860 - 2.866 & 1481 & 80 & 85 & 1404 & 2170 & 1404 & 1406 & 1404 \\
2.873 - 2.879 & 1306 & 70 & 73 & 1462 & 2213 & 1462 & 1465 & 1462 \\
2.886 - 2.892 & 1529 & 78 & 74 & 1393 & 2232 & 1393 & 1394 & 1393 \\
2.899 - 2.905 & 1420 & 88 & 94 & 1447 & 2269 & 1447 & 1448 & 1447 \\
2.912 - 2.918 & 1477 & 75 & 74 & 1449 & 2282 & 1449 & 1450 & 1449 \\
2.925 - 2.931 & 1457 & 70 & 70 & 1471 & 2326 & 1471 & 1472 & 1471 \\
2.938 - 2.944 & 1407 & 67 & 66 & 1468 & 2351 & 1468 & 1469 & 1468 \\
2.951 - 2.957 & 1542 & 72 & 72 & 1483 & 2363 & 1483 & 1484 & 1483 \\
2.964 - 2.970 & 1513 & 83 & 85 & 1488 & 2392 & 1488 & 1489 & 1488 \\
2.977 - 2.983 & 1537 & 82 & 85 & 1538 & 2417 & 1538 & 1539 & 1538 \\
2.990 - 2.996 & 1499 & 66 & 67 & 1527 & 2446 & 1527 & 1527 & 1527 \\
3.003 - 3.009 & 1524 & 71 & 74 & 1517 & 2467 & 1517 & 1518 & 1517 \\
3.016 - 3.022 & 1559 & 91 & 92 & 1588 & 2495 & 1588 & 1589 & 1588 \\
3.029 - 3.035 & 1546 & 62 & 64 & 1590 & 2540 & 1590 & 1590 & 1590 \\
3.042 - 3.048 & 1490 & 57 & 61 & 1545 & 2539 & 1545 & 1545 & 1545 \\
3.055 - 3.061 & 1637 & 70 & 75 & 1666 & 2555 & 1666 & 1666 & 1666 \\
3.068 - 3.074 & 1752 & 83 & 85 & 1594 & 2578 & 1594 & 1594 & 1594 \\
3.081 - 3.087 & 1600 & 65 & 67 & 1664 & 2612 & 1664 & 1664 & 1664 \\
3.094 - 3.100 & 1660 & 67 & 67 & 1657 & 2629 & 1657 & 1657 & 1657 \\
3.107 - 3.113 & 1667 & 63 & 67 & 1690 & 2672 & 1690 & 1690 & 1690 \\
3.120 - 3.126 & 1694 & 67 & 70 & 1722 & 2660 & 1722 & 1723 & 1722 \\
3.133 - 3.139 & 1763 & 70 & 74 & 1745 & 2682 & 1745 & 1745 & 1745 \\
3.146 - 3.152 & 1825 & 70 & 72 & 1778 & 2723 & 1778 & 1778 & 1778 \\
3.159 - 3.165 & 1800 & 71 & 73 & 1832 & 2726 & 1832 & 1832 & 1832 \\
3.172 - 3.178 & 1715 & 95 & 90 & 1770 & 2755 & 1770 & 1770 & 1770 \\
3.185 - 3.191 & 1898 & 68 & 70 & 1838 & 2754 & 1838 & 1838 & 1838 \\
3.198 - 3.204 & 1761 & 68 & 68 & 1768 & 2807 & 1768 & 1769 & 1768 \\
3.211 - 3.217 & 1791 & 69 & 69 & 1948 & 2813 & 1948 & 1948 & 1948 \\
3.224 - 3.230 & 1900 & 78 & 76 & 1853 & 2812 & 1853 & 1853 & 1853 \\
3.237 - 3.243 & 1907 & 62 & 69 & 1876 & 2842 & 1876 & 1876 & 1876 \\
3.250 - 3.256 & 1993 & 65 & 67 & 1964 & 2844 & 1964 & 1965 & 1964 \\
3.263 - 3.269 & 2012 & 89 & 91 & 1992 & 2860 & 1992 & 1993 & 1992 \\
3.276 - 3.282 & 2152 & 73 & 74 & 1948 & 2884 & 1948 & 1948 & 1948 \\
3.289 - 3.295 & 1956 & 69 & 73 & 1998 & 2952 & 1998 & 1998 & 1998 \\
3.302 - 3.309 & 2048 & 72 & 71 & 2047 & 2917 & 2047 & 2048 & 2047 \\
3.316 - 3.322 & 2112 & 76 & 75 & 2158 & 3008 & 2158 & 2159 & 2158 \\
3.329 - 3.335 & 2190 & 72 & 73 & 2193 & 2926 & 2193 & 2194 & 2193 \\
3.342 - 3.348 & 2052 & 72 & 74 & 2095 & 2944 & 2095 & 2096 & 2095 \\
3.355 - 3.361 & 2123 & 66 & 67 & 2190 & 2956 & 2190 & 2191 & 2190 \\
3.368 - 3.374 & 2255 & 84 & 88 & 2181 & 3004 & 2181 & 2182 & 2181 \\
3.381 - 3.387 & 2261 & 72 & 75 & 2191 & 2999 & 2191 & 2192 & 2191 \\
3.394 - 3.400 & 2323 & 76 & 76 & 2364 & 3147 & 2364 & 2366 & 2364 \\
3.407 - 3.413 & 2267 & 64 & 68 & 2256 & 3045 & 2256 & 2258 & 2256 \\
3.420 - 3.426 & 2270 & 67 & 73 & 2292 & 3052 & 2292 & 2294 & 2292 \\
3.433 - 3.439 & 2454 & 78 & 79 & 2345 & 3057 & 2345 & 2349 & 2345 \\
3.446 - 3.452 & 2278 & 73 & 71 & 2380 & 3077 & 2380 & 2383 & 2380 \\
3.459 - 3.465 & 2312 & 72 & 70 & 2346 & 3072 & 2346 & 2350 & 2346 \\
3.472 - 3.478 & 2449 & 78 & 78 & 2354 & 3083 & 2354 & 2360 & 2354 \\
3.485 - 3.491 & 2386 & 78 & 73 & 2418 & 3088 & 2418 & 2429 & 2418 \\
3.498 - 3.504 & 2522 & 78 & 78 & 2419 & 3104 & 2419 & 2427 & 2419 \\
3.511 - 3.517 & 2374 & 67 & 77 & 2447 & 3105 & 2447 & 2465 & 2447 \\
3.524 - 3.530 & 2486 & 88 & 86 & 2359 & 3118 & 2359 & 2371 & 2359 \\
3.537 - 3.543 & 2471 & 73 & 75 & 2472 & 3135 & 2472 & 2489 & 2472 \\
3.550 - 3.556 & 2471 & 75 & 78 & 2489 & 3136 & 2489 & 2502 & 2489 \\
3.563 - 3.569 & 2534 & 83 & 87 & 2523 & 3151 & 2523 & 2538 & 2523 \\
3.576 - 3.582 & 2575 & 71 & 74 & 2530 & 3149 & 2530 & 2552 & 2530 \\
3.589 - 3.595 & 2586 & 75 & 74 & 2638 & 3152 & 2638 & 2666 & 2638 \\
3.602 - 3.608 & 2546 & 84 & 79 & 2647 & 3156 & 2647 & 2669 & 2647 \\
3.615 - 3.621 & 2537 & 69 & 71 & 2646 & 3162 & 2646 & 2672 & 2646 \\
3.628 - 3.634 & 2659 & 77 & 86 & 2691 & 3194 & 2691 & 2719 & 2691 \\
3.641 - 3.647 & 2820 & 88 & 85 & 2735 & 3217 & 2735 & 2756 & 2735 \\
3.654 - 3.660 & 2589 & 65 & 70 & 2708 & 3236 & 2708 & 2734 & 2708 \\
3.667 - 3.673 & 2703 & 77 & 78 & 2771 & 3233 & 2771 & 2794 & 2771 \\
3.680 - 3.686 & 2814 & 67 & 68 & 2782 & 3263 & 2782 & 2810 & 2782 \\
3.693 - 3.699 & 2752 & 80 & 80 & 2809 & 3274 & 2809 & 2834 & 2809 \\
3.706 - 3.712 & 2838 & 85 & 79 & 2774 & 3234 & 2774 & 2805 & 2774 \\
3.831 - 3.839 & 2914 & 69 & 73 & 2983 & 3365 & 2983 & 3003 & 2983 \\
3.848 - 3.856 & 2983 & 71 & 74 & 3027 & 3366 & 3027 & 3050 & 3027 \\
3.865 - 3.873 & 3109 & 81 & 82 & 3021 & 3394 & 3021 & 3041 & 3021 \\
3.882 - 3.890 & 3208 & 84 & 82 & 3135 & 3406 & 3135 & 3160 & 3135 \\
3.899 - 3.907 & 3220 & 78 & 85 & 3114 & 3405 & 3114 & 3134 & 3114 \\
3.915 - 3.923 & 3005 & 77 & 79 & 3111 & 3409 & 3111 & 3135 & 3111 \\
3.932 - 3.940 & 3131 & 76 & 80 & 3152 & 3426 & 3152 & 3173 & 3152 \\
3.949 - 3.957 & 3200 & 86 & 86 & 3130 & 3465 & 3130 & 3149 & 3130 \\
3.966 - 3.974 & 2939 & 81 & 85 & 3068 & 3436 & 3068 & 3089 & 3068 \\
3.983 - 3.991 & 3101 & 72 & 77 & 3120 & 3482 & 3120 & 3138 & 3120 \\
4.000 - 4.008 & 3077 & 82 & 86 & 3045 & 3463 & 3045 & 3062 & 3045 \\
4.016 - 4.024 & 3077 & 88 & 92 & 3096 & 3477 & 3096 & 3114 & 3096 \\
4.033 - 4.041 & 3007 & 85 & 94 & 3082 & 3491 & 3082 & 3095 & 3082 \\
4.050 - 4.058 & 3061 & 83 & 92 & 3078 & 3526 & 3078 & 3094 & 3078 \\
4.067 - 4.075 & 3133 & 92 & 88 & 3146 & 3616 & 3146 & 3158 & 3146 \\
4.084 - 4.092 & 3107 & 97 & 98 & 2936 & 3540 & 2936 & 2948 & 2936 \\
4.100 - 4.108 & 2993 & 88 & 89 & 3023 & 3537 & 3023 & 3037 & 3023 \\
4.117 - 4.125 & 2992 & 90 & 93 & 3010 & 3536 & 3010 & 3019 & 3010 \\
4.134 - 4.142 & 2975 & 79 & 82 & 3032 & 3578 & 3032 & 3040 & 3032 \\
4.151 - 4.159 & 2955 & 87 & 92 & 2989 & 3563 & 2989 & 2996 & 2989 \\
4.168 - 4.176 & 2900 & 107 & 163 & 2972 & 3516 & 2967 & 2973 & 2971 \\
4.185 - 4.193 & 2987 & 98 & 97 & 2992 & 3600 & 2992 & 3000 & 2993 \\
4.201 - 4.209 & 3088 & 100 & 183 & 3029 & 3663 & 3045 & 3079 & 3057 \\
4.218 - 4.226 & 2928 & 94 & 92 & 2998 & 3583 & 2998 & 3004 & 3004 \\
4.235 - 4.243 & 2835 & 95 & 98 & 2938 & 3590 & 2938 & 2942 & 2945 \\
4.252 - 4.260 & 2983 & 101 & 99 & 2947 & 3598 & 2941 & 2956 & 2952 \\
4.269 - 4.277 & 2860 & 95 & 95 & 2903 & 3602 & 2903 & 2907 & 2913 \\
4.286 - 4.294 & 2983 & 146 & 107 & 2992 & 3652 & 2932 & 2990 & 3075 \\
4.302 - 4.310 & 2876 & 108 & 183 & 3031 & 3604 & 3062 & 3029 & 3026 \\
4.319 - 4.327 & 2905 & 109 & 116 & 3059 & 3569 & 3156 & 3067 & 3006 \\
4.336 - 4.344 & 2816 & 106 & 110 & 3035 & 3496 & 3163 & 3090 & 3018 \\
4.353 - 4.361 & 3009 & 101 & 137 & 3005 & 3374 & 3114 & 3031 & 3000 \\
4.370 - 4.378 & 2925 & 112 & 155 & 2733 & 2813 & 3135 & 2767 & 2777 \\
4.387 - 4.395 & 2885 & 113 & 129 & 2900 & 3100 & 3160 & 2994 & 2981 \\
4.403 - 4.411 & 3061 & 109 & 167 & 2925 & 3017 & 3348 & 2989 & 2913 \\
4.420 - 4.428 & 2834 & 103 & 136 & 2914 & 3294 & 3170 & 2955 & 2927 \\
4.437 - 4.445 & 2898 & 114 & 173 & 2849 & 3105 & 3060 & 2822 & 2871 \\
4.454 - 4.462 & 2799 & 108 & 185 & 2915 & 3132 & 3127 & 2973 & 2965 \\
4.471 - 4.479 & 2800 & 120 & 182 & 2771 & 2814 & 3216 & 2793 & 2769 \\
4.487 - 4.495 & 2807 & 119 & 182 & 2832 & 2924 & 3174 & 2830 & 2860 \\
4.504 - 4.512 & 3008 & 123 & 115 & 2775 & 2995 & 3109 & 2768 & 2705 \\
4.521 - 4.529 & 2757 & 117 & 124 & 3026 & 3219 & 3338 & 3032 & 3023 \\
4.538 - 4.546 & 3115 & 115 & 169 & 2852 & 2985 & 3367 & 2885 & 2850 \\
4.555 - 4.563 & 2875 & 118 & 197 & 3007 & 3274 & 3397 & 3067 & 3062 \\
4.572 - 4.580 & 3025 & 123 & 128 & 2857 & 3039 & 3299 & 2840 & 2806 \\
4.588 - 4.596 & 3071 & 168 & 184 & 2977 & 3153 & 3314 & 2965 & 2997 \\
4.605 - 4.613 & 2914 & 129 & 163 & 3249 & 3461 & 3604 & 3258 & 3223 \\
4.622 - 4.630 & 3105 & 125 & 185 & 2917 & 3017 & 3360 & 2994 & 2973 \\
4.639 - 4.647 & 3101 & 120 & 173 & 3002 & 3170 & 3343 & 3087 & 3093 \\
4.656 - 4.664 & 3137 & 133 & 161 & 3113 & 3425 & 3279 & 3197 & 3100 \\
4.673 - 4.681 & 3146 & 142 & 113 & 2990 & 3317 & 3151 & 2987 & 2949 \\
4.689 - 4.697 & 3143 & 133 & 164 & 3217 & 3548 & 3359 & 3292 & 3279 \\
4.706 - 4.714 & 3086 & 130 & 144 & 3075 & 3402 & 3122 & 3038 & 3035 \\
4.723 - 4.731 & 3273 & 135 & 175 & 3052 & 3564 & 3111 & 3043 & 3036 \\
4.740 - 4.748 & 3205 & 134 & 197 & 3092 & 3389 & 3375 & 3098 & 3027 \\
4.757 - 4.765 & 2978 & 137 & 144 & 3110 & 3453 & 3373 & 3151 & 3139 \\
4.773 - 4.781 & 3160 & 136 & 100 & 3115 & 3524 & 3298 & 3165 & 3145 \\
4.790 - 4.798 & 3121 & 147 & 107 & 3076 & 3407 & 3253 & 3029 & 3015 \\
4.807 - 4.815 & 3063 & 136 & 155 & 2967 & 3191 & 3252 & 2943 & 2963 \\
4.824 - 4.832 & 3326 & 148 & 196 & 3261 & 3608 & 3451 & 3261 & 3227 \\
4.841 - 4.849 & 3004 & 142 & 122 & 3095 & 3425 & 3291 & 3057 & 3086 \\
4.858 - 4.866 & 3341 & 156 & 147 & 3198 & 3547 & 3380 & 3157 & 3111 \\
4.874 - 4.882 & 3090 & 149 & 115 & 3113 & 3497 & 3220 & 3106 & 3125 \\
4.891 - 4.899 & 3128 & 193 & 250 & 3045 & 3371 & 3142 & 3041 & 3035 \\
4.908 - 4.916 & 3203 & 157 & 106 & 3160 & 3420 & 3321 & 3104 & 3147 \\
4.925 - 4.933 & 2978 & 143 & 100 & 3097 & 3362 & 3283 & 3034 & 3013 \\
4.942 - 4.950 & 3005 & 144 & 108 & 2988 & 3254 & 3167 & 2960 & 2971 \\
4.959 - 4.967 & 3100 & 148 & 139 & 3001 & 3173 & 3277 & 3038 & 3011 \\
4.975 - 4.983 & 3286 & 139 & 162 & 3235 & 3565 & 3357 & 3291 & 3207 \\
4.992 - 5.000 & 3028 & 167 & 188 & 3112 & 3522 & 3258 & 3141 & 3120 \\
5.009 - 5.017 & 3252 & 153 & 191 & 3154 & 3300 & 3201 & 3148 & 3125 \\
5.026 - 5.034 & 3061 & 156 & 106 & 3006 & 3273 & 3100 & 3093 & 3033 \\
5.043 - 5.051 & 3368 & 155 & 194 & 3219 & 3876 & 3217 & 3277 & 3255 \\
5.060 - 5.068 & 3085 & 169 & 164 & 3102 & 3609 & 3214 & 3117 & 3175 \\
5.076 - 5.084 & 3170 & 156 & 108 & 3112 & 3688 & 3154 & 3122 & 3124 \\
5.093 - 5.101 & 3290 & 144 & 111 & 3226 & 3861 & 3273 & 3250 & 3273 \\
5.110 - 5.118 & 3077 & 166 & 157 & 3195 & 3880 & 3259 & 3181 & 3164 \\
5.127 - 5.135 & 3082 & 162 & 118 & 3205 & 3808 & 3295 & 3214 & 3232 \\
5.144 - 5.152 & 3050 & 152 & 196 & 3069 & 3339 & 3150 & 3024 & 3068 \\
5.160 - 5.168 & 3130 & 160 & 150 & 3058 & 3345 & 3131 & 3039 & 3004 \\
\caption{Summary of the spectral data shown on Figure~\ref{fig:spectrum}.}
\label{tab:spectraldata}\\
\end{longtable}



\bibliography{references}{}
\bibliographystyle{aasjournal}



\end{document}